\documentclass[12pt,aps,prd,superscriptaddress,nofootinbib, preprintnumbers]{revtex4-1}

\usepackage[english]{babel}
\usepackage[latin1]{inputenc}
\usepackage{hyperref}
\usepackage{graphicx,color}
\usepackage{latexsym}
\usepackage{amsmath}
\usepackage{amsfonts,dsfont}
\usepackage{amssymb,slashed}
\usepackage{tensor}
\usepackage{bbold}
\usepackage{verbatim,epigraph}
\usepackage{xcolor}
\usepackage{soul}

\begin{document}

\title{Equations of Motion as Covariant Gauss Law: The Maxwell-Chern-Simons Case}

\author{\textbf{A. P. Balachandran}\footnote{balachandran38@gmail.com} }

\affiliation{Physics Department, Syracuse University, Syracuse, New York, U.S.A.}

\author{\textbf{Arshad Momen}\footnote{arshad@iub.edu.bd, amomen@du.ac.bd}}  

\affiliation{Department of Physical Sciences, 
Independent University,
Bashundhara R/A, Dhaka-1212, Bangladesh}

\altaffiliation{On leave of absence from Theoretical Physics Department, 
University of Dhaka, Dhaka, Bangladesh. }

\author{\textbf{Amilcar R. de Queiroz}\footnote{amilcarq@unb.br}}  

\affiliation{{Instituto de Fisica, Universidade de
Brasilia,} \\ \emph{Caixa Postal 04455, 70919-970, Brasilia, DF, Brazil}}


\begin{abstract}
Time-independent gauge transformations are implemented in the canonical formalism by 
the Gauss law which is not covariant. The covariant form of Gauss law is conceptually important 
for studying asymptotic properties of the gauge fields. For QED in $3+1$ 
dimensions, we have developed a formalism for treating the equations of motion (EOM) 
themselves as constraints, that is, constraints on  states using Peierls' quantization 
\cite{ABLM}. They generate spacetime dependent gauge transformations. We extend 
these results to the Maxwell-Chern-Simons (MCS) Lagrangian. The surprising result is that 
the covariant Gauss law commutes with all observables: the gauge invariance of the 
Lagrangian gets trivialized upon quantization. The calculations do not fix a gauge. We also 
consider a novel gauge condition on test functions (not on quantum fields) which we name 
the ``quasi-self-dual gauge'' condition. It explicitly shows the mass spectrum of the 
theory. In this version, no freedom remains for the gauge transformations: EOM 
commute with all observables and are in the center of the algebra of observables.
\end{abstract}

\maketitle

\section{Introduction}

The Abelian Maxwell-Chern-Simons (MCS) theory \cite{Dunne:1998qy} is a theory of a 
massive ``photon'' in $2+1$ dimensions. It violates parity, $P$, and time-reversal, $T$. 
The Lagrangian has $U(1)$ gauge invariance, but it is absent in the final Hamiltonian.

Our focus is on the fate of this $U(1)$ gauge group. We will see that it has a trivial 
action on the connection potentials $A_\mu$ after covariant quantization and that the 
operator which generates them is the operator which implements EOM by vanishing on 
quantum states \cite{ABLM}: the gauge symmetry of the Lagrangian disappears on 
quantization.

This approach which does not impose gauge conditions on $A_\mu$ will be contrasted with 
an alternative approach which is also new and does not fix the gauge of $A_\mu$. It is 
covariant and quickly shows why $A_\mu$ has mass. It is not $P$ and $T$ invariant and 
also does not lead to EOM as constraints which generates gauge transformations. The EOM 
are actually in the center of the algebra of observables in both of these approaches.

We interpret EOM as generalized covariantized Gauss laws. This is reasonable: a component 
of the Maxwell equation, say $\partial^\mu F_{\mu 0}=0$, for the field strength $F$ is in 
fact the Gauss law. The collection of such component Gauss laws with regard to every 
Cauchy surface and their superpositions give the field equations. This justifies our 
assumptions.

Let\footnote{Many of the equations were supplied to A.P.B. by V.P. Nair.} $A_\mu$ be a 
vector field in $2+1$ dimensions and consider the action, with $\eta_{\mu \nu}= 
\text{diag}(-,+,+)$,
\begin{align}\label{MCS-action-1}
 S&=\int d^3x~\Big(-\frac{1}{4} F_{\mu \nu}F^{\mu \nu} + \frac{k e^2}{4\pi} 
\varepsilon^{\mu \nu \sigma} A_\mu \partial_\nu A_\sigma \Big), \\
 F_{\mu \nu}&= \partial_\mu A_\nu - \partial_\nu A_\mu.
\end{align}
It gives the equations of motion (EOM)
\begin{equation}\label{EOM-MCS-1}
 \partial^\mu F_{\mu \nu} + m F_\nu = 0, \qquad m\equiv \frac{ke^2}{2\pi},
\end{equation}
where
\begin{equation}\label{Dual-Field-1}
 F_\nu = \varepsilon_{\nu \mu \sigma} \partial^\mu A^\sigma.
\end{equation}
Notice that
\begin{equation}\label{Div-Dual-Field}
 \partial^\mu F_\mu = 0.
\end{equation}
Writing $F^{\mu \nu}$ in terms fo $F^\mu$, we obtain
\begin{equation}\label{EOM-MCS-2}
 -\varepsilon_{\mu \nu \sigma} \partial^\nu F^\sigma + m F_\mu = 0.
\end{equation}
Applying $\varepsilon^{\mu \lambda \rho}\partial_\rho$, using (\ref{Div-Dual-Field}) and 
also (\ref{EOM-MCS-2}) to eliminate the term with $\varepsilon_{\mu \nu \sigma}$, we 
obtain
\begin{equation}\label{Massive-EOM-1}
 \Box F_\mu + m^2 F_\mu = 0, \qquad \qquad \Box\equiv \partial_\nu \partial^\nu,
\end{equation}
so that, as it is well-known, MCS describes a massive photon $F_\mu$.

\section{The Causal Commutator}

We give the causal commutator in the Lorentz gauge,
\begin{equation}\label{Lorentz-gauge-1}
 \partial^\mu A_\mu = 0,
\end{equation}
so that 
\begin{equation}\label{EOM-Lorentz-gauge-1}
 \Box A_\mu + m \varepsilon_{\mu \nu \sigma} \partial^\nu A^\sigma = 0.
\end{equation}
Then, the causal commutator is
\begin{align}\label{Causal-Commutator-1}
 D_{\mu \nu}(x-y) \equiv \left[A_\mu(x),A_\nu(y) \right] = \int_{\mathcal{C}} 
~\frac{d^3 p}{(2\pi)^3}~e^{-ip\cdot(x-y)} M_{\mu \nu}(p),
\end{align}
where
\begin{align}
M^{\mu \nu}(p) = \frac{i}{p^2-m^2}~\left( \eta^{\mu \nu} - \frac{p^\mu p^\nu}{p^2}- i m 
\varepsilon^{\sigma \mu \nu} \frac{p_\sigma}{p^2} \right),
\end{align}
and
\begin{equation}\label{Div-D-1}
 \partial_\mu D^{\mu \nu}(x-y)=0.
\end{equation}
The contour $\mathcal{C}$ encloses the poles at 
\begin{equation}\label{Poles-1}
p_0=\pm\sqrt{\vec{p}^{~2}+m^2}.
\end{equation}

The novelty in this paper is that we will not use the gauge condition 
(\ref{Causal-Commutator-1})below. Rather we work with field $A(\eta)$ smeared with smooth 
test-functions $\eta_\mu$ which are compactly supported,
\begin{align}
 A(\eta) &= \int d^3x~\eta^{\mu}(x) A_\mu(x), \qquad \qquad \eta^\mu \in 
\mathcal{C}^\infty_0(\mathds{R}^3), \\
\partial^\mu \eta_\mu &= 0.\label{condition} 
\end{align}
Here the zero subscript denotes compact support and infinity infinite differentiability.
As Roepstorff \cite{Roepstorff} has discussed, (\ref{Div-D-1}) is gauge invariant by 
partial integration,
\begin{equation}
 \big(A+\partial \Lambda\big)(\eta) = A(\eta), \qquad \text{for} \qquad \Lambda \in 
\mathcal{C}^\infty_0(\mathds{R}^3).
\end{equation}
The algebra of $A(\eta)$ is inferred from (\ref{Causal-Commutator-1}) as
\begin{align}\label{algebra-obs-1}
 \big[A(\eta_1),A(\eta_2)\big] = \int d^3x d^3y ~ (\eta_1)^\mu(x)~D_{\mu 
\nu}(x-y)~(\eta_2)^\nu(y),
\end{align}
with $\eta_i^\mu\in \mathcal{C}^\infty_0(\mathds{R}^3)$ and $\partial_\mu \eta^\mu_i=0$, 
$i=1,2$. 

The algebra with commutator (\ref{algebra-obs-1}) defined by the local observables 
$A(\eta)$ defines MCS. It involves 
no gauge fixing of $A$.

\section{EOM as Constraints}

The classical equations of motion are (\ref{EOM-MCS-1}). We smear them with test function 
$\rho^\mu\in\mathcal{C}^\infty(\mathds{R})$ and write them as an equation involving no 
derivatives of $A_\mu$. This is appropriate since we should write derivatives of 
distribution $A_\mu$ at the quantum level as derivatives of test functions $\rho^\mu$. 

Let us introduce the notations
\begin{align}
 \hat{F}_{\mu\nu}(\rho) &= \partial_\mu \rho_\nu - \partial_\nu \rho_\mu, \\
 \hat{F}_{\mu\nu}(\rho)(x) &= \partial_\mu \rho_\nu(x) - \partial_\nu \rho_\mu(x).
\end{align}

We do not insist on requiring  $\partial_\mu \rho^\mu=0$. 
Multiplying (\ref{EOM-MCS-1}) by $\rho^\nu$ and integrating, we obtain classically the 
equations
\begin{equation}
 G[\rho]\equiv \int \Big(\partial^\mu \hat{F}_{\mu \sigma}(\rho) + m \varepsilon_{\sigma \mu 
\nu} \partial^\mu \rho^\nu \Big) A^\sigma=0.
\end{equation}
We regard the LHS at the quantum level as an operator $G[\rho]$ which vanishes on allowed 
quantum states:
\begin{equation}\label{G-action-2}
 G[\rho]~|\cdot\rangle = \int d^3x~\Big(\partial^\mu \hat{F}_{\mu \sigma} + m 
\varepsilon_{\sigma \mu \nu} \partial^\mu \rho^\nu \Big) A^\sigma (x)~ |\cdot\rangle = 0.
\end{equation}
This defines the domain of the observables $A(\eta)$. 

Note that even though $\rho^\mu$ does not fulfill Lorentz gauge, the function in 
(\ref{G-action-2}) multiplying $A^\sigma$ does and is a proper test function for 
$A^\sigma$.

We must show that $G[\rho]$'s are first class. That result follows below.

\section{The Commutator $[G[\rho],A_\sigma(y)]$}

We find that the commutator $[G[\rho],A_\beta(y)]$ is identically zero. It implies that 
$G[\rho]$'s commute for different $\rho$ and hence are first class constraints.

We have
\begin{equation}\label{Commutator-GA-1}
\big[G[\rho],A_\sigma(y)\big]=\int d^3x~\Big(\partial_\mu \hat{F}^{\mu \kappa}[\rho]+m 
\varepsilon^{\kappa \mu \nu} \partial_\mu \rho_\nu \Big)~D_{\kappa \sigma}(x-y).
\end{equation}
Now,
\begin{equation}
 \partial_\mu \hat{F}^{\mu \kappa}[\rho] = \Box \rho^\kappa - \partial^\kappa\big(\partial\cdot 
\rho \big),
\end{equation}
and then under the integration the second term $\partial^\kappa\big(\partial\cdot 
\rho \big)$ vanishes due to partial integration 
and use of $\partial^\mu D_{\mu \sigma}=0$. 

As for the remaining terms, we can write (\ref{Commutator-GA-1}) as
\begin{align}
 \big[G[\rho],A_\sigma(y)\big]&=\int d^3x \Big(\Box_x\rho^\kappa + m \varepsilon^{\kappa \mu 
\nu} \partial_\mu \rho_\nu  \Big) ~ \big[A_\kappa (x),A_\sigma(y) \big] \nonumber \\
                              &= \int d^3 x~\Big[\Big(\Box_x\rho^\kappa + m 
\varepsilon^{\kappa \mu 
\nu} \partial_\mu \rho_\nu  \Big) A_\kappa(x) , A_\sigma(y) \Big],
\end{align}
where the subscript $x$ means differentiation with respect to $x$. Now, this expression
vanishes after integration by parts and use of (\ref{EOM-Lorentz-gauge-1}). Therefore,
\begin{equation}\label{Commutator-GA-2}
 \big[G[\rho],A_\sigma(y)\big]=0.
\end{equation}

\section{A Novel Gauge Condition}

The test function $\rho$ (unlike $\eta^\mu$ satisfying $\partial_\mu \eta^\mu=0$) is not 
so far subjected to any gauge condition. Nor is $A_\mu$. We will now ``gauge fix'' the 
test functions by imposing the condition
\begin{equation}\label{TF-gauge-fix-1}
 \rho_\mu = m \varepsilon_{\mu \nu \sigma} \partial^\nu \rho^\sigma.
\end{equation}
It implies that $\rho_\mu$ itself is transverse,
\begin{equation}
 \partial^\mu \rho_\mu=0.
\end{equation}
The condition (\ref{TF-gauge-fix-1}) is not gauge invariant and hence is a gauge fixing 
condition. 

From the condition (\ref{TF-gauge-fix-1}) we have the following facts:
\begin{enumerate}
 \item\label{F1} The result that $A_\mu$ describes massive vector bosons becomes explicit;
 \item\label{F2} The EOM \emph{commutes} as before with all $A_\mu$ and does not generate 
gauge transformations. It is in the centre of the algebra of observables.
\end{enumerate}

As for item (\ref{F1}), we can look at (\ref{G-action-2}) and impose 
(\ref{TF-gauge-fix-1}). That gives
\begin{equation}
 G[\rho]~|\cdot\rangle = \int d^3x \Big(\Box \rho^\kappa + m^2 \rho^\kappa \Big)~ 
A_\kappa~|\cdot\rangle = 0,
\end{equation}
which on partial integration gives the result
\begin{equation}
 \big(\Box + m^2\big)A_\rho = 0,
\end{equation}
classically. So $A_\rho$ has mass $m$.

As for item (\ref{F2}), (\ref{Commutator-GA-2}) is true for any choice of $\rho$, hence 
the fact follows.

\section{Significance}

The link between EOM and gauge transformations seems significant. It has not been 
discussed previously prior to \cite{ABLM}. It has now turned up in QED, linearised 
gravity \cite{Balachandran-LG} and Maxwell-Chern-Simons theory. This link is of course 
present in $U(1)$ gauge theories in all dimensions.

In non-abelian gauge theories, like QCD, the commutator of the fields $A_\mu$ at distinct 
points $x$ and $y$ is not known due to the non-linearity of the field equations. We are 
hence not able to analyse this case in the present framework.

\section{Further Problems}

In Dirac's approach to constrained dynamics, one distinguishes between the  first and second class 
constraints. Often gauge fixing conditions are introduced to turn the former into the 
second class. Second class constraints can be eliminated using Dirac-Bergmann brackets 
\cite{Balachandran:2017jha}. All of these happen on a Cauchy hypersurface, that is, at a fixed time.

In this paper, we have introduced EOM as first class constraints. It is natural to ask: 
Is there an analogous theory of constraints in this spacetime picture? This appears to be 
an open interesting problem.

\end{document}